\documentstyle[amsmath,amssymb,graphicx]{article}

\def\be{\begin{eqnarray}}
\def\ee{\end{eqnarray}}
\def\nn{\nonumber}

\def\Tr{{\rm Tr}\,}

\def\l[{\phantom.[}

\hoffset= - 1.0in         

\begin{document}

\hfill ITEP/TH-21/16

\hfill IITP/TH-15/16

\bigskip

\centerline{\Large{
Rectangular superpolynomials for the figure-eight knot
}}

\bigskip

\bigskip

\centerline{{\bf Ya.Kononov$^{d,e}$, A.Morozov$^{a,b,c}$}}

\bigskip

{\footnotesize

\centerline{{\small
$^a$ ITEP, Moscow 117218, Russia}}

\centerline{{\small
$^b$ National Research Nuclear University MEPhI, Moscow 115409, Russia
}}

\centerline{{\small
$^c$ Institute for Information Transmission Problems, Moscow 127994, Russia
}}

\centerline{{\small
$^d$ Higher School of Economics, Math Department, Moscow, 117312, Russia
}}

\centerline{{\small
$^e$ Landau Institute for Theoretical Physics, Chernogolovka, Russia
}}
}

\bigskip

\bigskip

\centerline{ABSTRACT}

\bigskip

{\footnotesize
We rewrite the recently proposed differential expansion formula
for HOMFLY polynomials of the knot $4_1$ in arbitrary rectangular
representation $R=[r^s]$ as a sum over all Young sub-diagrams
$\lambda$ of $R$
with extraordinary simple coefficients $D_{\lambda^{tr}}(r)\cdot D_\lambda(s)$
in front of the $Z$-factors.
Somewhat miraculously, these coefficients are made from quantum dimensions
of symmetric representations of the groups $SL(r)$ and $SL(s)$
and restrict summation to diagrams with no more than $s$ rows and $r$ columns.
They possess a natural $\beta$-deformation to Macdonald dimensions
and produces {\it positive polynomials}, which can be considered as plausible
candidates for the role of the rectangular superpolynomials.
Both polynomiality and positivity are non-evident properties of arising expressions,
still they are true.
This extends the previous suggestions for symmetric
and antisymmetric representations (when $s=1$ or $r=1$ respectively) to arbitrary
rectangular representations.
As usual for differential expansion, there are additional gradings.
In the only  example, available for comparison -- that of the trefoil knot $3_1$,
to which our results for $4_1$ are straightforwardly extended, -- one of them reproduces
the "fourth grading" for hyperpolynomials.
Factorization properties are nicely preserved even in the 5-graded case.
}

\bigskip

\bigskip

\section{Introduction}

Superpolynomials are among the main mysteries of modern theoretical physics.
There is a lot of evidence, that they exist, but there is neither a clear conceptual definition
nor a clear algorithm for practical evaluation -- despite many efforts during the last
two decades.
This paper reviews the problem from the most naive direction and reports some small
progress, based on the recent development in adjacent fields.

The story begins from a mysterious discovery
that Wilson loop averages
\be
{\cal H}_R^{\cal K}(A,q) = \left<\Tr_R P\exp\oint_{\cal K} {\cal A} \right>
\label{Wilsonline}
\ee
in $3d$ Chern-Simons theory \cite{CS}, which depend only on the topology of the embedding
${\cal K}\hookrightarrow R^3$, i.e. on the topology of the {\it knot} ${\cal K}$,
are Laurent polynomials with integer coefficients -- when expressed in terms of peculiar variables
$q=\exp\left(\frac{2\pi i}{k+N}\right)$ and $A=q^N$ instead of the coupling constant $k$
and rank of the gauge group $SU(N)$.
This fact explains the name {\it knot polynomials} \cite{knotpols} for ${\cal H}_R$, its various
reductions at special values of $A$ and $q$ and generalizations to other gauge groups.
$H_R$ {\it per se} is also known as  HOMFLY polynomial, {\it colored} by a Young diagram $R$,
which labels irreducible finite-dimensional representations of $SU(N)$.
Integrality property implies that the coefficients of knot polynomials count something --
these can be numbers of certain physical states in some stringy models \cite{GSV}, underlying the
Chern-Simons theory,
while in an abstract categorification program
{\it something} means just
dimensions of {\it some} vector spaces.
Though $K$-theory allows "dimensions" to be negative, it would be simpler to have them positive --
and this was the reason to look for generalization ({\it refinement}) of $H_R$,
which should have all the coefficients positive.
Convention is that this {\it super}polynomial
${\cal P}_R^{\cal K}(A,q,T)$ has all coefficients positive integers and satisfies the reduction property
\be
{\cal P}_R^{\cal K}(A,q,T=-1)={\cal H}_R^{\cal K}(A,q)
\ee
Knot-dependent complexes for which ${\cal P}$ and $H$ are respectively the Poincare and Euler polynomials
are explicitly constructed for the single-box diagram $R=[1]=\Box$ at fixed $N$ \cite{Kho,KhR},
but calculations are technically involved for $N>2$
and there are only restricted successes for other representations $R$.
An intriguing possibility is to consider superpolynomial as a $p$-adic generalization of HOMFLY \cite{padic}.
In fact the number of additional arguments can be larger than two,
$q \longrightarrow \{q,-qT\} \longrightarrow \{q_{i}\}$
are often thought as relatives to parameters of Kerov polynomials, and at least three-parametric
deformations are already considered, both in the context of knot polynomials \cite{GGS,arthdiff}
and for the closely related DIM-algebras \cite{Ok,DIM}.
The literature on superpolynomials and their physical interpretation is rich,
see, for example, \cite{DGR}-\cite{NO}.

\bigskip

If there is any commonly accepted definition today,
then it describes the superpolynomial as

1) a Laurent polynomial ${\cal P}_R^{\cal K}({\bf a},{\bf q},{\bf t})$ with all coefficients
positive integers,

2) reproducing HOMFLY at ${\bf t}=-1$,

3) reproducing Khovanov-Jones polynomial at ${\bf a}={\bf q}^2$

4) reproducing Khovanov-Rozansky at ${\bf a} = {\bf q}^N$

\bigskip

The problem with this "definition" is that Khovanov's is an explicit,
but {\it ad hoc} construction, which is largely technical
and not quite related with the "best" available conceptual
definitions of HOMFLY -- either from functional integrals and quantum-group theory
or from various recursions.
While HOMFLY are defined (at least mnemonically) in (\ref{Wilsonline})
as $3d$ topological invariants,
Khovanov's construction is in terms of knot diagrams --
the knot's projections on $2d$ planes, with topological invariance
substituted by Reidemeister invariance, which is natural for representations of braids,
but is only respected,
but not fully explicit, in Khovanov's approach.
Khovanov's complex is built with the help of a hypercube of colorings \cite{Kho,hcube,Anosup},
and direct relation to the quantum-${\cal R}$-matrix Reshetikhin-Turaev
formalism \cite{RT,inds,RTmod} is available only at $N=2$, where a peculiar Kauffman's
${\cal R}$-matrix \cite{KaufR} can be used.
Khovanov-Rozansky generalization to $N>2$ involves
sophisticated matrix-factorization technique, which is very hard to use
in practical calculations.
Partly for this reason the construction of colored superpolynomials
is still unclear, especially in non-rectangular representations.

Thus it is not a big surprise that there is a continuous search for a
simpler and more straightforward definitions of superpolynomials
and/or reformulations of HOMFLY calculus, which would allow to describe
superpolynomials as natural and unambiguous deformations.

At advanced level the story is about refined Chern-Simons and, more generally,
refined topological strings --
which is now linked to a deformation from loop Kac-Moody to toroidal
Ding-Iohara-Miki algebras \cite{DIM}.
However, applications  to knot theory are still work in progress.
The real issue here is that knot theory in appropriate formulation
is almost identical to representation theory, but is in fact a little more stable:
knot theory can survive deformations which break the representation theory structures.
The two recent examples are:

(i) applicability of Vogel's universality \cite{Vog} to knots
(knot invariants in the $E_8$-sector contain Vieta-like combinations of the roots of certain universal
polynomials, rather than the roots themselves, which represent particular quantum
dimensions -- and thus possess simple universal expressions, even when dimensions
fail to do so \cite{unipols}), and

(ii) pretzel knot polynomials are expressed through bilinear combinations of Racah matrices
($6j$-symbols) \cite{mmmspre}, and these possess nice $\beta$-deformations \cite{ArthSha}, while
this is not true for the $6j$-symbols themselves.

Precise identification of this "healthy" part of representation theory,
which in practice is captured by consideration of knot polynomials,
remains a puzzling open question -- and this is a probable reason for persisting
problems with adequate definition and evaluation of super- and hyper-polynomials.

\bigskip

At naive level attempts are made to deform HOMFLY into superpolynomials
for particular families of knots.
If there is any {\it canonical} construction of this type,
applicable universally to {\it all} knots,
it should reveal and take into account the representation dependence
of HOMFLY, because in any particular representation superpolynomials
are believed to involve more knot invariants than HOMFLY,
and balance is restored (if at all) only for the entire set of polynomials
in all representations.

In this paper we report new results in the most promising approach
of this kind --
the one based on the differential expansions \cite{DGR,IMMMfe,evo,arthdiff,arthlinks,Konodef}.
The point is that colored HOMFLY can be re-expanded in powers of
the {\it differentials} $D_n=\{Aq^n\} = Aq^n - \frac{1}{Aq^n}$,
and these expansions have more pronounced representation dependence
than original polynomials.
Moreover, the differentials $D_n$ are directly related to differentials
in Khovanov-Rozansky complexes \cite{DGR},
and the formalism naturally simplifies transition to superpolynomials.
The method is most adequate for a family of twisted knots
(see below), where it allowed to fully describe HOMFLY and superpolynomials
 in arbitrary symmetric
representation and derive various recursions in $R$,
which serve as a model for more complicated knots and links.
Complexity of the differential expansion is regulated by
a {\it defect} \cite{Konodef}, which is actually a degree in $q^2$ of
the fundamental Alexander polynomial in topological framing minus one.
What distinguishes twisted knots is that they have defect zero
and the differential expansion is
actually strengthened to one in $Z$-factors, which are bilinear
combinations $Z_{n|m}=D_nD_{-m}=\{Aq^n\}\{A/q^m\}$,
and have a very clear superpolynomial
deformation to ${\cal Z}_{n|m}=\{Aq^n\}\{A/t^m\}$.

In this paper we use the recently discovered
HOMFLY polynomials in all rectangular representations and their differential expansion
for the figure-eight knot $4_1$  \cite{rect41}
to conjecture the corresponding rectangular twisted superpolynomials,
which are in accord with various previous speculations.
Also straightforward in this case is the next, forth grading \cite{GGS},
which seems to be naturally built into the differential-expansion
formalism \cite{arthdiff}.

Comparison with alternative approaches, involving $c$-factors \cite{DMMSS},
DAHA algebra (a part of DIM) \cite{Che,CheDa,Berest}, Verlinde algebras \cite{AgSha},
$T$-deformed ${\cal R}$-matrices \cite{AnoM,GaMoore} etc
is very desirable, but such strong results are not yet available there.
Hopefully this paper will stimulate new progress in these other directions.
Also of interest are generalizations to other defect-zero knots.
Really challenging remain the cases of non-vanishing defects and especially
of non-rectangular representations, where the structure of differential expansion
remain obscure and sometime questioned is the very existence of superpolynomials
(see \cite{Ano21,MMMS21,mmms1} for partial description of the simplest $R=[21]$ and
$R=[31]$ cases).
Last, but not the least, reduction of superpolynomials to finite $N$ and the difference
between reduced and unreduced superpolynomials are not yet captured
by differential expansion tools -- therefore in this paper we deal only with
reduced superpolynomials at generic $A$.

\bigskip

The main results of this paper are eqs.(\ref{41rsmod}) and (\ref{41rssup}).

The first is a strong and inspiring reformulation of the recent result of \cite{rect41},
and it is based on an amusing decomposition formula (\ref{binodeco}) for binomial
coefficients, specific for rectangular representations.

The second is immediate, though non-trivial, lift to superpolynomials.

In addition to this we discuss two extra gradings (one of them -- related to that of
\cite{GGS} and \cite{arthdiff}), which are implied preserve the nice factorization
properties of the polynomials and can deserve further investigation.








\section{Reinterpretation of the formula for rectangular HOMFLY}

Differential expansion can be considered as a $q$-deformation of the formula
for special polynomials \cite{DMMSS,spepo}, arising from reduced HOMFLY in the limit $q=1$:
for any knot ${\cal K}$ and in any representation $R$
\be
H^{\cal K}_R(q=1,A)\ = \ \Big(H^{\cal K}_{_\Box}(q=1,A)\Big)^{|R|}
\ee
Moreover, in topological framing reduced HOMFLY turns into unity, when $A=q^{\pm 1}$,
what means that
\be
H^{\cal K}_{_\Box}(q,A) = 1 + F^{\cal K}_{_\Box} (q,A)\cdot\{Aq\}\{A/q\}
\ee
with $\{x\} := x-x^{-1}$ and some function $F^{\cal K}_{_\Box} (q,A)$,
which for the figure-eight knot ${\cal K}=4_1$ is just unity.
It follows that the special polynomial
\be
H^{{4_1}}_R(q=1,A) \ = \ \Big(1 +  \{A\}^2 \Big)^{|R|} =
\sum_{k=0}^{|R|}  \binom{|R|}{k} \{A\}^{2k}
\label{spedeco}
\ee
Differential expansion \cite{IMMMfe,evo,arthdiff,Konodef} substitutes the $q$-independent powers $\{A\}^{2k}$
and binomial coefficients
by more involved representation-dependent $k$-linear combinations of $Z$-factors
$Z_{i|j} = \{Aq^i\}\{A/q^j\}$ with $q$-dependent coefficients.
The structure of this deformation strongly depends on the "defect" of the knot \cite{Konodef},
which is miraculously regulated by the power of Alexander polynomial $H_{_\Box}(q,A=1)$,
arising from HOMFLY at $A=1$, and is not fully revealed yet.

The biggest achievement at this moment is
the recently suggested in  formula \cite{rect41} for the differential expansion
of rectangular HOMFLY polynomials for $4_1$
\be
H_{[r^s]}^{4_1} = \sum_{F=0}^{{\rm min}(r,s)} \sum_{
\stackrel{0\leq a_F <\ldots < a_3<a_2<a_1<r}{0\leq b_F < \ldots < b_3< b_2<b_1<s}}
\ \ \prod_{f'<f''}^F
\left( \frac{[a_{f'}-a_{f''}] [b_{f'}-b_{f''}]}{[a_{f'}+b_{f''}+1] [a_{f''}+b_{f'}+1]} \right)^2 \cdot
\nn
\ee
\vspace{-0.35cm}
\be
\cdot \prod_{f=1}^F \left( h^{a_f + b_f + 1}
\left(\frac{[a_f+b_f]!}{([a_f]![b_f]! }\right)^2
\frac{[r+b_f]![s+a_f]!}{[r-1-a_f]![s-1-b_f]!\big([a_f+b_f+1]!\big)^2}
 \underline{\prod_{i_f=-b_f}^{a_f} \{Aq^{r+i_f}\}\{Aq^{i_f-s}\}}\right)
\label{41rs}
\ee
and its further deformation to arbitrary twisted knots in \cite{rectwist}.
For convenience we introduced here the forth-grading \cite{GGS,arthdiff} parameter $h$,
which counts the number of $Z$-factors -- actually, for HOMFLY $h=1$.
Quantum numbers are defined as $[x] = \frac{\{q^x\}}{\{q\}}= \frac{q^x-q^{-x}}{q-q^{-1}}$.

Our primary task in this paper is to reveal the structure of this complicated expression and rewrite
it in a very simple form (\ref{41rsmod}).

\bigskip

The first observation is that the sum in (\ref{41rs}) is actually over Young diagrams $\lambda$,
which are sub-diagrams of rectangular $R=[r^s]$:
\be
H_{[r^s]}(4_1) \ \stackrel{(\ref{41rs})}{=} \
\sum_{\lambda\subset [r^s]} h^{|\lambda|} \cdot {\cal C}_\lambda^{[r^s]}(q)\cdot {Z}^\lambda_{r|s}(A,q)
\label{41rsinterm}
\ee
where the $A$-dependent $Z^\lambda_{r|s}$, underlined in (\ref{41rs}), is  a product of "shifted" $Z$-factors
\be
{ Z}^\lambda_{r|s}(A,q) = \prod_{\square \in \lambda} Z_{r|s}^{(a'(\square) - l'(\square))} =
\prod_{\square \in \lambda} \{A q^{r + a'(\square) - l'(\square)}\} \{A  q^{-s + a'(\square) - l'(\square)}\}
\label{Zlambda}
\ee

The second observation is that the $A$-independent coefficients ${\cal C}_\lambda$ have a very simple form.
To understand this it is necessary first to return to (\ref{spedeco}) and decompose binomial coefficients
into contributions of Young sub-diagrams.
Namely, for factorized $|R|=rs$ there is a remarkable decomposition:
\be
\boxed{
\binom{rs}{k} = \frac{(rs)!}{k!\,(rs-k)!}= \sum_{|\lambda| = k} d_{\lambda^{tr}} (r)\cdot d_{\lambda} (s)
}
\label{binodeco}
\ee
where
\be
\!\!\!\!\!\!\!\!\!\!\!\!\!\!\!\!\!\!\!\!\!\!\!\!\!\!\!\!\!\!\!\!\!\!\!\!\!\!\!\!\!\!\!\!\!\!\!\!\!\!\!\!
\!\!\!\!\!\!\!\!\!\!\!\!
d_\lambda (N)  = \prod_{\square \in \lambda}
\frac{N - l'(\square) + a'(\square)}{a(\square) + l(\square) + 1}
\ee
\begin{picture}(120,25)(-300,15)
\put(20,100){\line(1,0){100}}
\put(20,80){\line(1,0){100}}
\put(20,60){\line(1,0){80}}
\put(20,40){\line(1,0){40}}
\put(20,20){\line(1,0){20}}
\put(20,100){\line(0,-1){80}}
\put(40,100){\line(0,-1){80}}
\put(60,100){\line(0,-1){60}}
\put(80,100){\line(0,-1){40}}
\put(100,100){\line(0,-1){40}}
\put(120,100){\line(0,-1){20}}

\put(40,70){\vector(-1,0){20}}
\put(50,80){\vector(0,1){20}}
\put(60,70){\vector(1,0){40}}
\put(50,60){\vector(0,-1){20}}

\put(10,67){\mbox{$a'$}}
\put(48,105){\mbox{$l'$}}
\put(105,67){\mbox{$a$}}
\put(48,30){\mbox{$l$}}
\end{picture}

\noindent
is dimension of representations $\lambda$ of the Lie algebra $sl_N$,
while $a,l,a',l'$ are arms, legs, co-arms and co-legs, associated with a box $\square$
inside the Young diagram.
The simplest particular cases of (\ref{binodeco}) are:
$$
\binom{rs}{2} = \binom{r+1}{2} \binom{s}{2} + \binom{r}{2} \binom{s+1}{2}
= d_{[2]}(r)\cdot d_{[11]}(s)+d_{[11]}(r)\cdot d_{[2]}(s)
$$
$$
\binom{rs}{3} = \binom{r+2}{3} \binom{s}{3} + 4\cdot \binom{r+1}{3}\binom{s+1}{3}
+ \binom{r}{3} \binom{s+2}{3}
= d_{[3]}(r)\cdot d_{[111]}(s) + d_{[21]}(r)\cdot d_{[21]}(s) + d_{[111]}(r)\cdot d_{[3]}(s)
$$
$$
\ldots
$$
and it is easy to check that at $q=1$ the sophisticated coefficients in (\ref{41rs}) are given
by exactly these formulas.

Decomposition (\ref{binodeco}) follows from the Cauchy identity in the form
\be
\exp\left(\sum_{k=1}^\infty \frac{(-h )^k p_k\bar p_k}{k}\right) =
\sum_\lambda h^{|\lambda|}\cdot{\rm Schur}_{\lambda^{tr}}\{p\}\cdot {\rm Schur}_{\lambda}\{\bar p\}
\ee
for all $p_k = r$ and $\bar p_k=s$.
Indeed, for this choice of time-variables
\be
{\rm Schur}_\lambda\{{\rm all}\ p_k=r\} = d_\lambda(r)
\label{cladimSchur}
\ee
and  the identity reduces to
\be
(1+h)^{rs} = \sum_\lambda h^{|\lambda|}\cdot d_{\lambda^{tr}}(r)\cdot d_{\lambda}(s)
\ee
It is now clear what the $q$-deformation is: the quantum version of (\ref{cladimSchur}) is
\be
S_\lambda^*(N|q)={\rm Schur}_\lambda\{ p_k=p_k^*\} = D_\lambda(N)
\ee
where the time-variables are restricted to topological locus
\be
p_k^* = \frac{\{A^k\}}{\{q^k\}} \ \stackrel{A=q^N}{=} \ \frac{[Nk]}{[k]}
\ee
and $D_\lambda(N)$ are quantum dimensions of representation $\lambda$ of the quantum group
$U_q(sl_N)$:
\be
D_\lambda(N) = \prod_{\square \in \lambda}
\frac{[N - l'(\square) + a'(\square)]}{[a(\square) + l(\square) + 1]}
\ee

Surprisingly or not, this quantum deformation provides the $q$-dependent coefficients in
(\ref{41rs}) and it acquires the very explicit and simple form (\ref{41rsinterm}):
\be
\boxed{\boxed{
H_{[r^s]}^{4_1}(q,A) \ \stackrel{(\ref{41rs})}{=} \
\sum_{\lambda\subset [r^s]} h^{|\lambda|} \cdot D_{\lambda^{tr}}(r)\cdot D_\lambda(s)
\cdot {Z}^{\lambda}_{r|s}(A,q)
}}
\label{41rsmod}
\ee
with $Z$-factors (\ref{Zlambda}).

Since $D_{\lambda^{tr}}(N|q) = D_\lambda(N,q^{-1})$,
while ${Z}^{\lambda^{tr}}_{r|s}(A,q)={Z}^{\lambda}_{s|r}(A,q^{-1})$,
we get the usual symmetry property
$H_{[r^s]}(q,A) = H_{[s^r]}(q^{-1},A)$.
Since $D_\lambda(r)$ vanishes for all $\lambda$ with more than $r$ columns, the sum in (\ref{41rsmod})
is automatically restricted to $\lambda$ with no more than $r$ columns and $s$ rows, i.e. to
sub-diagrams of the original $R=[r^s]$.

\bigskip

In somewhat more abstract language the Cauchy identity is the well-known decomposition of $GL(V)\times GL(W)$-module
\be
\Lambda^*_h (V\otimes W) = \sum_\lambda h^{|\lambda|} S_\lambda V \otimes S_{\lambda^t} W,
\ee
and the classical dimensions are just
\be
d_\lambda (N) = \dim S_\lambda ({\mathbb C}^N)
\ee
and if we view
\be
x = q^r, \ \ \ y = q^s
\ee
as equivariant parameters and $\ \{x\} = x-x^{-1}\ $ -- as the K-theoretical A-genus \cite{Ok},  then
\be
H^{4_1}_{[r^s]} \ \stackrel{(\ref{41rsmod})}{=} \ 1 + \sum_{\lambda} h^{|\lambda|} \prod_{\square \in \lambda}
\frac{\{x q^{a'_\square-l'_\square}\}\{y q^{l'_\square-a'_\square}\}}{\{q^{a_\square+l_\square+1}\}\{q^{a_\square+l_\square+1}\}}
\{A x q^{l'_\square - a'_\square}\} \{A y^{-1} q^{  l'_\square - a'_\square}\}
\ee
is actually the
Lefshetz fixed-point formula \cite{lfp},
applied to a certain sheaf on the Hilbert scheme (fixed points of the torus action are labeled by Young diagrams).
This means that all rectangular HOMFLY polynomials can be read out of the $(q,q)$-equivariant
Euler characteristic of a certain sheaf (universal for all representations).

\bigskip

Now it is natural to consider $(q,t)$-equivariant characteristics, i.e. to relax the condition $q = t$
to $t=q^\beta$.

\section{$\beta$-deformation}

The universal part of the $\beta$-deformation
of differential expansions is the change of the $Z$-factors:
\begin{multline}
Z_{r|s}^{(i|j)} = \{Aq^{r+i-j}\}\{Aq^{-s+i-j}\} \ \ \ \ \ \longrightarrow \ \ \ \ \
{\cal Z}_{r|s}^{(i|j)} = \{Aq^{r+i}/t^j\}\{Aq^i/t^{s+j}\}
=\\
= (-)^{r+1}\cdot\left({\bf a}^{2}  {\bf q}^{r-s+2\,i-2\,j} {\bf t}^{r+2\,i+1}  +   {\bf q}^{r+s}{\bf t}^r
+ \frac{1}{ {\bf q}^{r+s}{\bf t}^r} + \frac{1}{{\bf a}^{2}  {\bf q}^{r-s+2\,i-2\,j} {\bf t}^{r+2\,i+1}}\right)
\label{defrule}
\end{multline}
which,
modulo an overall sign, is a positive Laurent polynomial in the DGR variables \cite{DGR}
\be
{\bf q} = t \ \ \ \ \ \ {\bf t}= -q/t \ \ \ \ \ \ \ {\bf a} = A \sqrt{t/q},
\label{bfvariables}
\ee
For composite $Z$-factors (\ref{Zlambda}) the shifts $i$ and $j$ are well defined
as  $a'$ and $l'$ respectively, and (\ref{defrule}) implies
\be
{ Z}^\lambda_{r|s}(q,A) \to  {\cal Z}^\lambda_{r|s}(q,t,A) =
\prod_{\square \in \lambda} {\cal Z}_{r|s}^{(a'_{_\square}|l'_{_\square})} =
\prod_{\square \in \lambda} \{  A q^{r + a'_{_\square}} /t^{  l'_{_\square}} \}
\{A  q^{a'_{_\square}}/ t^{  s + l'_{_\square}}\}
\label{calZlambda}
\ee

Challenging is the deformation of the coefficients in front of the $Z$-factors.
For figure-eight knot the suggestion of \cite{IMMMfe} was to rewrite the formula for
symmetric HOMFLY in the form, where all the coefficients are just unities:
\be
H^{4_1}_{[r]}(q,A) = 1 + \sum_{k=1}^r \frac{[r]!}{[k]![r-k]!} \prod_{i=0}^{k-1} \{Aq^{r+1}\}\{Aq^{i-1}\} =
\sum_{k=0}^r \  \sum_{1 \leq i_1 < \ldots < i_k \leq r}
Z_{i_1} (A) Z_{i_2} (Aq) \ldots Z_{i_k} (A q^{k-1}),
\ee
where in this case the relevant $Z$-factors  are
\be
Z_i (A) = \{Aq^{2(r-i)+1}\}\{A/q\}
\ee
Since $r-i\geq 0$ they are naturally deformed $Z_i(A) \to  {\cal Z}_i(A) = \{Aq^{2(p-i)+1}\}\{A/t\}$,
and the deformation (symmetric superpolynomial) is just
\be
\!\!\!
P^{4_1}_{[r]}(q,t,A)  =
\sum_{k=0}^r \   \sum_{1 \leq i_1 < \ldots < i_k \leq r} \!\!\!\!\!
{\cal Z}_{i_1} (A){\cal Z}_{i_2} (Aq) \ldots {\cal Z}_{i_k} (A q^{k-1})
= 1 + \sum_{k=1}^r \underbrace{\frac{[r]_q!}{[k]_q![r-k]_q!}}_{M^{t^{-1},q^{-1}}_{[1^k]}(A=q^r)}
\prod_{i=0}^{k-1} \{Aq^{r+i}\}\{Aq^{i}/t\}
\label{Psym}
\ee
i.e. the binomial coefficients actually remain intact -- the ratios of $q$-numbers,
while $t$ appears only in the ${\cal Z}$.
Conversely, for antisymmetric representations binomial coefficients are made fully from the $t$-numbers:
\be
P^{4_1}_{[1^s]}(q,t,A)
= 1 + \sum_{k=1}^s \underbrace{\frac{[s]_t!}{[k]_t![s-k]_t!}}_{M^{q,t}_{[1^k]}(A=t^s)}
\prod_{i=0}^{k-1} \{A/t^{s+1}\}\{Aq/t^{i}\}
\label{Pasym}
\ee
In these formulas we showed also, that the binomial coefficients
are expressible through Macdonald dimensions $M^{q,t}_R(A)$ -- the values of
Macdonald polynomials $M_R^{q,t}\{p\}$ at the topological locus $p_k=p_k^*(A,t):=\frac{ \{A^k\} }{\{t^k\}}$,
$$
M_{[1]}^{q,t}(A) = \frac{\{A\}}{\{t\}}
$$
$$
M_{[2]}^{q,t}(A) = \frac{\{A\}\{Aq\}}{\{t\}\{qt\}} \ \ \ \ \ \ \ \
M_{[11]}^{q,t}(A) = \frac{\{A\}\{A/t\}}{\{t\}\{t^2\}}
$$

$$
M_{[3]}^{q,t}(A) = \frac{\{A\}\{Aq\}\{Aq^2\}}{\{t\}\{qt\}\{q^2t\}}\ \ \ \ \ \ \
M_{[21]}^{q,t}(A) = \frac{\{A\}\{Aq\}\{A/t\}}{\{t\}^2\{qt^2\}} \ \ \ \ \ \ \
M_{[111]}^{q,t}(A) = \frac{\{A\}\{A/t\}\{A/t^2\}}{\{t\}\{t^2\}\{t^3\}}
$$

$$
M_{[4]}^{q,t}(A) = \frac{\{A\}\{Aq\}\{Aq^2\}\{Aq^3\}}{\{t\}\{qt\}\{q^2t\}\{q^3t\}} \ \ \ \ \ \ \
M_{[31]}^{q,t}(A) = \frac{\{A\}\{Aq\}\{Aq^2\}\{A/t\}}{\{t\}^2\{qt\}\{q^2t^2\}} \ \ \ \ \ \
M_{[22]}^{q,t}(A) = \frac{\{A\}\{Aq\}\{A/t\}\{Aq/t\}}{\{t\}\{t^2\}\{qt\}\{qt^2\}}
$$
\be
M_{[211]}^{q,t}(A) = \frac{\{A\}\{Aq\}\{A/t\}\{A/t^2\}}{\{t\}^2\{t^2\}\{qt^3\}} \ \ \ \ \ \ \
M_{[1111]}^{q,t}(A) = \frac{\{A\}\{A/t\}\{A/t^2\}\{A/t^3\}}{\{t\}\{t^2\}\{t^3\}\{t^4\}} \nn \\
 \ldots
 \label{Mdims}
\ee
Note that sums in both expressions (\ref{Psym}) and (\ref{Pasym})
for symmetric and antisymmetric $R=[r]$ and $R=[1^s]$  involve only single-row??? diagrams $[1^k]$,
but differ by the change $(q,t)\longrightarrow (t^{-1},q^{-1})$, which also applies to the values of $p_k^*$ --
the change is performed directly in (\ref{Mdims}).
It also deserves noting that this change is the usual ingredient
of the transposition rule for Macdonald polynomials
\be
M_{\lambda^{tr}}^{q,t}\{p_k\} =
M_{\lambda}^{t^{-1},q^{-1}}\!\!\left(-\frac{\{t^k\}}{\{q^k\}}p_k\right)
\cdot \prod_{\square \in \lambda}
\left(- \frac{\{q^{l_\square} t^{a_\square + 1}\}}{\{q^{l_\square + 1} t^{a_\square}\}} \right)
\ee
which substitutes the simple one for Schur functions,
\be
{\rm Schur}_{\lambda^{tr}}\{p_k\} = (-)^{|\lambda|}\cdot {\rm Schur}_{\lambda}\{-p_k\}
\ \ \ \ \ \Longrightarrow \ \
D_{\lambda^{tr}}(N|q) = D_{\lambda^{tr}}(N|q^{-1}) = (-)^{|\lambda|}\cdot D_{\lambda}(-N|q)
\ee

\bigskip

Combining this observation with the new expression (\ref{41rsmod}) for  combinatorial coefficients,
one can easily guess the $\beta$-deformation of all rectangular HOMFLY polynomials.
In abbreviated notation the suggestion is
\be
\boxed{\boxed{
P^{4_1}_{[r^s]}(q,t,A) =
\sum_{\lambda\subset [r]^s} h^{|\lambda|}  \cdot
{\cal M}^{tr}_{\lambda^{tr}}(r)\cdot {\cal M}_{\lambda}(s) \cdot
{\cal Z}_{r|s}^\lambda
}}
\label{41rssup}
\ee
with the ${\cal Z}$-factor from the r.h.s. of (\ref{calZlambda}).
More explicitly,
\be
P^{4_1}_{[r^s]}(q,t,A)
= \sum_{\lambda} h^{|\lambda|}  \cdot
\overbrace{M^{t^{-1},q^{-1}}_{\lambda^{tr}}\!\!\left(A=q^r\right)}^{{\cal M}^{tr}_{\lambda^{tr}}(r)}
\cdot  \overbrace{M^{q, t}_{\lambda}\left(A=t^s\right)}^{{\cal M}_{\lambda}(s)}
\cdot {\cal Z}_{r|s}^\lambda
= \nn \\
= \sum_{k=1}^p h^{|\lambda|}  \cdot
M^{t^{-1},q^{-1}}_{\lambda^{tr}}\!\!\left(p_i = \frac{\{q^{-ri}\}}{\{q^{-i}\}}\right)
\cdot  M^{q, t}_{\lambda} \!\left(p_i = \frac{\{t^{si}\}}{\{t^i\}}\right)
\cdot {\cal Z}_{r|s}^\lambda
= \nn \\
= \ 1 + \sum_{\lambda\subset [r^s]} h^{|\lambda|} \underbrace{\prod_{\square \in \lambda}
\frac{\{q^{r-a'_{_\square}} t^{l'_{_\square}}\}}{\{t^{l_\square} q^{a_{_\square}+1}\}}
\frac{\{t^{s-l'_{_\square}} q^{a'_{_\square}}\}}{\{t^{l_{_\square}+1} q^{a_{_\square}}\}}
\{A q^{r + a'_{_\square}} t^ { - l'_{\square}}\} \{A  q^{  a'_{_\square}} t^{-s  - l'_{_\square}}\}
}_{contr_{_\lambda}}
\label{rectsup}
\ee
The sign of the ${\cal Z}$ factor, originating from $(-)^{r+1}$ in (\ref{defrule}),
is compensated by exactly the same sign, arising in ${\cal M}_{\lambda^{tr}}^{tr}(r)$
after the change (\ref{bfvariables}).
A more serious problem could be that, in variance with quantum dimensions $D_\lambda(N|q)$,
Macdonald dimensions ${\cal M}_\lambda(N|q)$
are {\it not} Laurent polynomials, even for concrete integer values of $N$.
Surprisingly or not, however, the numerators disappear after summation over
all sub-diagrams $\lambda$ (actually, they do so in every order in $|\lambda|$),
and (\ref{rectsup}) {\bf   is always (Laurent) polynomial and positive in the DGR variables}
 (\ref{bfvariables})!
Eq.(\ref{rectsup}) reproduces all previously suggested formulas for colored super- and hyper- polynomials
of $4_1$ (and also $3_1$),
with the single exception of that in \cite{ArthSha}
(which, however, deviates already from the conventional answer for the fundamental representation with $r=s=1$).

\section{Polynomiality \label{Pol}}

To demonstrate how polynomiality emerges it deserves providing a couple of examples.

\bigskip

In the case of representation $R=[2]$ contributing are just three sub-diagrams
\be
contr_{_{[ \ \ ]}} = 1
\nn
\ee
\begin{multline}
contr_{_{[1]}} = [2]_q \{Aq^2\}\{A/t\}
   \to {\bf a}^2 {\bf q}^2 {\bf t}^4+\frac{1}{{\bf a}^2 {\bf q}^2 {\bf t}^4}+{\bf a}^2 {\bf t}^2+\frac{1}{{\bf a}^2 {\bf t}^2}+{\bf q}^4 {\bf t}^3+\frac{1}{{\bf q}^4 {\bf t}^3}+{\bf q}^2
   {\bf t}+\frac{1}{{\bf q}^2 {\bf t}}
   \nn
\end{multline}
\begin{multline}
contr_{_{[2]}} = \{Aq^2\}\{A/t\}\{Aq^3\}\{Aq/t\} \to \\
   \to {\bf a}^4 {\bf q}^4 {\bf t}^8+\frac{1}{{\bf a}^4 {\bf q}^4 {\bf t}^8}+{\bf a}^2 {\bf q}^6 {\bf t}^7+\frac{1}{{\bf a}^2 {\bf q}^6 {\bf t}^7}+{\bf a}^2 {\bf q}^4
   {\bf t}^5+\frac{1}{{\bf a}^2 {\bf q}^4 {\bf t}^5}+\frac{{\bf a}^2 {\bf t}}{{\bf q}^2}+\frac{{\bf q}^2}{{\bf a}^2 {\bf t}}+{\bf a}^2 {\bf t}^3+\frac{1}{{\bf a}^2
   {\bf t}^3}+{\bf q}^6 {\bf t}^4+\frac{1}{{\bf q}^6 {\bf t}^4}+{\bf q}^2 {\bf t}^2
   +\frac{1}{{\bf q}^2 {\bf t}^2}+2
   \nn
\end{multline}
There is exactly one diagram in each order $|\lambda|$, thus each of the tree contributions
is {\it per se} a positive polynomial.

\bigskip

In the case of representation $R=[2,2]$ the number of contributing  sub-diagrams is already six:
\be
contr_{_{[ \ \ ]}} = 1
\nn
\ee
{\small
\begin{multline}
contr_{_{[1]}} = [2]_q [2]_t \{Aq^2\}\{A/t^2\}\to \\
   \to
   {\bf a}^2 {\bf q}^2 {\bf t}^4+\frac{1}{{\bf a}^2 {\bf q}^2 {\bf t}^4}+\frac{{\bf q}^2}{{\bf a}^2 {\bf t}^2}+\frac{{\bf a}^2 {\bf t}^2}{{\bf q}^2}+{\bf a}^2
   {\bf t}^4+\frac{1}{{\bf a}^2 {\bf t}^4}+{\bf a}^2 {\bf t}^2+\frac{1}{{\bf a}^2 {\bf t}^2}+{\bf q}^6 {\bf t}^3+\frac{1}{{\bf q}^6 {\bf t}^3}+{\bf q}^4
   {\bf t}^3+\frac{1}{{\bf q}^4 {\bf t}^3}+{\bf q}^4 {\bf t}+\frac{1}{{\bf q}^4 {\bf t}}+{\bf q}^2 {\bf t}+\frac{1}{{\bf q}^2 {\bf t}}
\nn\end{multline}
\begin{multline}
contr_{_{[2]}} = [2]_t \frac{\{qt^2\}}{\{qt\}}{\{Aq^3\}\{Aq^2\}\{Aq/t^2\}\{A/t^2\}}\to \\
\frac{\left({\bf q}^2+1\right) \left({\bf q}^3 {\bf t}-1\right) \left({\bf q}^3 {\bf t}+1\right) \left({\bf a}^2 {\bf t}+{\bf q}^4\right)
   \left({\bf a}^2 {\bf t}^3+{\bf q}^2\right) \left({\bf a}^2 {\bf q}^4 {\bf t}^5+1\right) \left({\bf a}^2 {\bf q}^6 {\bf t}^7+1\right)}{{\bf a}^4
   {\bf q}^{10} {\bf t}^8 \left({\bf q}^2 {\bf t}-1\right) \left({\bf q}^2 {\bf t}+1\right)}
\nn\end{multline}
\begin{multline}
contr_{_{[1,1]}} = [2]_q \frac{\{q^2t\}}{\{qt\}}{\{Aq^2\}\{Aq^2/t\}\{A/t^2\}\{A/t^3\}}\to \\
   \frac{\left({\bf q}^2 {\bf t}^2+1\right) \left({\bf q}^3 {\bf t}^2-1\right) \left({\bf q}^3 {\bf t}^2+1\right) \left({\bf a}^2
   {\bf t}+{\bf q}^4\right) \left({\bf a}^2 {\bf t}+{\bf q}^6\right) \left({\bf a}^2 {\bf q}^2 {\bf t}^5+1\right) \left({\bf a}^2 {\bf q}^4
   {\bf t}^5+1\right)}{{\bf a}^4 {\bf q}^{10} {\bf t}^8 \left({\bf q}^2 {\bf t}-1\right) \left({\bf q}^2 {\bf t}+1\right)}
\nn\end{multline}
\begin{multline}
contr_{_{[2,1]}} = [2]_q [2]_t \{Aq^2\}\{A/t^2\}\{Aq^3\}\{Aq/t^2\}\{Aq^2/t\}\{A/t^3\} \to
   \\
   \to \frac{\left({\bf q}^2+1\right) \left({\bf q}^2 {\bf t}^2+1\right) \left({\bf a}^2 {\bf t}+{\bf q}^4\right) \left({\bf a}^2
   {\bf t}+{\bf q}^6\right) \left({\bf a}^2 {\bf t}^3+{\bf q}^2\right) \left({\bf a}^2 {\bf q}^2 {\bf t}^5+1\right) \left({\bf a}^2 {\bf q}^4
   {\bf t}^5+1\right) \left({\bf a}^2 {\bf q}^6 {\bf t}^7+1\right)}{{\bf a}^6 {\bf q}^{14} {\bf t}^{12}}
\nn
\end{multline}
\begin{multline}
contr_{_{[2,2]}} = \{Aq^2\}\{A/t^2\}\{Aq^3\}\{Aq/t^2\}\{Aq^2/t\}\{A/t^3\}\{A*q^3/t\}\{Aq/t^3\}
   \to \\
   \to
   \frac{1}{{\bf a}^8 {\bf t}^{16} {\bf q}^{16}}\left({\bf a}^2 {\bf t}+{\bf q}^4\right) \left({\bf a}^2 {\bf t}+{\bf q}^6\right) \left({\bf a}^2 {\bf t}^3+{\bf q}^2\right) \left({\bf a}^2
   {\bf t}^3+{\bf q}^4\right) \left({\bf a}^2 {\bf q}^2 {\bf t}^5+1\right) \left({\bf a}^2 {\bf q}^4 {\bf t}^5+1\right) \left({\bf a}^2 {\bf q}^4
   {\bf t}^7+1\right) \left({\bf a}^2 {\bf q}^6 {\bf t}^7+1\right)
   \nn
\end{multline}
}
Two of them have the same size $\Big|[2]\Big|=\Big|[1,1]\Big|$, and  their individual contributions
have non-positive factors both in the denominators and the numerators.
However, when added, they provide a positive polynomial.
Moreover, they still produce a polynomial, if added with the coefficients $c_{[2]}$ and $c_{[11]}$,
provided
\be
c_{[2]}-c_{[11]} \sim \{qt\}
\label{c211}
\ee

\section{Plethystic logarithm of the series}

All rectangular figure-eight superpolynomials are made from a single series:
\begin{multline}
K^{4_1} := \ 1 + \sum_{\lambda} h^{|\lambda|} \prod_{\square \in \lambda}
\frac{\{xq^{-a'_{_\square}} t^{l'_{_\square}}\}}{\{t^l_{_\square} q^{a_{_\square}+1}\}}
\frac{\{yt^{-l'_{_\square}} q^{a'_{_\square}}\}}{\{t^{l_{_\square}+1} q^{a_{_\square}}\}}
\{A x q^{ a'_{_\square}} t^ { - l'_{_\square}}\} \{A  q^{  a'_{_\square}} y^{-1} t^{  - l'_{_\square}}\} = \\
=  1 + h \cdot\frac{\{x\}\{y\}}{\{q\}\{t\}} \{Ax\}\{A/y\} + h^2 \cdot \left(
\frac{\{x\}\{x/q\}\{y\}\{yq\}}{\{q\}\{q^2\}\{t\}\{qt\}} \{Ax\}\{Axq\}\{A/y\}\{Aq/y\} \right. +\\
\left. +
\frac{\{x\}\{xt\}\{y\}\{y/t\}}{\{t\}\{t^2\}\{q\}\{qt\}} \{Ax\}\{Ax/t\}\{A/y\}\{A/(yt)\}
\right) + O(h^3)
\end{multline}
Superpolynomials can be obtained by the specialization
\be
P_{[r^s]}^{4_1} = K\left(x = q^r, y=t^s \right)
\ee
It turns out that this series is a symmetric power of a more simple series $L$:
\be
K(x,y,A,q,t,h) = {Sym}^* (L) := \exp\left(\sum_{d=1}^\infty \ \frac{ L(x^d,y^d,A^d, q^d, t^d,  h^d)}{d}\right)
\ee
If one opens the brackets $\{\ldots\}$,
expansion of $K$ up to $h^3$ involves more than 900 items.
At the some time its plethystic logarithm $L$ is much simpler:

\be
L^{4_1} =    \frac{\{x\}\{y\}\{Ax\}\{A/y\}}{\{q\}\{t\}}\cdot \Big\{
h \ - \ h^2\cdot  \left(\alpha+\alpha^{-1}\right)  +
\ee
\vspace{-0.5cm}
\be
  \left.
+ h^3\cdot   \left(\alpha^2(q^2+t^{-2}-x^2-y^{-2}) - \frac{\alpha q}{t}(x^{-2}+y^2)
+1 - \frac{t}{\alpha q}(x^{2}+y^{-2})+\alpha^{-2}(q^{-2}+t^{2}-x^{-2}-y^{2})\right)
 - O(h^4)\right\}
\nn\ee
where $\alpha = A^2q/t$.
It would be interesting to find the meaning and the general term of this new expansion.

\section{Rectangular superpolynomials for the trefoil}

Figure-eight is the simplest representative of the family of twist knots,
which all have defect zero and thus possess a comparably simple differential expansion.
The difference between twist knots is that the contribution of each diagram $\lambda\subset R$ contains
an additional factor $F_\lambda$, which was found for the (anti)symmetric
representations in \cite{evo}, and extended to arbitrary rectangular HOMFLY
just recently in \cite{rectwist}.

The $\beta$-deformation of the $F$-factors is a separate problem, but in the
particular case of the trefoil $3_1$ -- another member of the twist knot --
family it is simple.
In this case the factors $F^{3_1}_\lambda$ and their $\beta$-deformations are
no more than simple monomials.
The answer is actually known for symmetric representations since \cite{DMMSS,IMMMfe} and \cite{evo}.
and generalization of our newly-discovered (\ref{41rssup}) to the case of $3_1$ is
\be
\boxed{
P^{3_1}_{[r^s]}(q,t,A) =
\sum_{\lambda\subset [r]^s} h^{|\lambda|}  \cdot
 \underbrace{\left(-A^2 q/t \right)^{|\lambda|} \left(
 \prod_{\square \in \lambda}
q^{2a_\square} t^{-2l_\square}
\right)}_{{\cal F}^{3_1}_\lambda}
\cdot
{\cal M}^{tr}_{\lambda^{tr}}(r)\cdot {\cal M}_{\lambda}(s) \cdot
{\cal Z}_{r|s}^\lambda
}
\label{31rssup}
\ee
Like in the case of $4_1$ this formula produces positive
Laurent polynomial -- despite particular items in the sum are
non-polynomial.
In the example of sec.\ref{Pol} this is because the condition
(\ref{c211}) is fulfilled by the factors $c_{[2]}={\cal F}^{3_1}_{[2]}$
and $c_{[11]}={\cal F}^{3_1}_{[11]}$.
Criteria of this kind can be used to define the $\beta$-deformation of the $F$-factors
from \cite{rectwist} for all other twist knots.

Important thing is that $3_1$ is not only twist, but also a torus knot,
thus this result can be compared with the torus hyperpolynomials
from \cite{Che} and their 4-grading generalizations \cite{GGS},
which for rectangular representations $R$ are believed to provide
the true superpolynomials (i.e. should coincide with the future
calculation in Khovanov's approach).
The result of this comparison is positive: (\ref{31rssup}) reproduces
Cherednik's polynomials \cite{Che} and their generalizations.
In Cherednik's case it is sufficient to substitute $A^2\longrightarrow -A^2$.
%
%
To reproduce quadruply graded knot homologies of \cite{GGS} for $3_1$ and $4_1$,
one inserts an additional  parameter $\sigma$ into the differentials,
i.e. further deforms the $Z$-factors, leaving the coefficients intact:
\be
\!\!\!
\boxed{
{\cal P}^{3_1}_{[r^s]}=\sum_{\lambda\subset [r]^s} h^{|\lambda|}  \cdot
 \underbrace{\left(-A^2 q/t \right)^{|\lambda|} \left(
 \prod_{\square \in \lambda}
q^{2a_\square} t^{-2l_\square}
\right)}_{{\cal F}^{3_1}_\lambda}
\cdot
{\cal M}^{tr}_{\lambda^{tr}}(r)\cdot {\cal M}_{\lambda}(s) \cdot
\prod_{\square \in \lambda} \{  A q^{r + a'_{_\square}} /\sigma  t^{  l'_{_\square}} \}
\{A  q^{a'_{_\square}} \sigma  / t^{  s + l'_{_\square}}\}
}
\label{31rssup4}
\ee
and then make the change of variables:
\be
\sigma  \to {\bf t}_r^{-s}, \ \ q \to -{\bf qt}_c, \ \  t \to {\bf q}/{\bf t}_r, \ \
A \to {\bf a} \sqrt{-{\bf t}_r {\bf t}_c}
\ee
(in these last formulas indices $r$ and $c$ are original notation of \cite{GGS}, this $r$ has nothing to do with
the diagram $R=r^s$, however, the exponent $s$ in the substitute of $\sigma$ {\it is} the number of columns
in $R$).
In the case of $3_1$ this extends the original suggestion of \cite{arthdiff} from (anti)symmetric to all
rectangular representations.
In the case of $4_1$ there are no formulas in \cite{GGS} beyond (anti)symmetric case,
thus (\ref{31rssup4}) with eliminated $F$-factor, ${\cal F}^{4_1}_\lambda=1$
is only a conjecture.

\section{Factorization properties}

Despite this is not {\it a priori} requested,
rectangular superpolynomial (\ref{rectsup}) has the following algebraic properties,
which generalize factorization rules for HOMFLY at roots of unity \cite{Anton,KM}:
\be
\begin{array}{lcccc}
{\rm at} \ q=1:   && P_{[r^s]} = \left( P_{[1^s]} \right)^r && {\rm for \ all} \ h, \  A \ {\rm and } \ t \\ \\
{\rm at} \ t=1:  &&  P_{[r^s]} =  \left( P_{[r]} \right)^s && {\rm for \ all} \  h, \ A \ {\rm and } \ q \\ \\
{\rm at} \ q^{2n}=1:   && P_{[r^s]} \cdot P_{[n^s]} = P_{[(r+n)^s]}&& {\rm for \ all} \  h, \ A \ {\rm and } \ t \\ \\
{\rm at} \ t^{2m}=1:  &&   P_{[r^s]} \cdot P_{[r^m]} = P_{[r^{s+m}]}&& {\rm for \ all} \ h, \ A \ {\rm and } \ q
\end{array}
\ee
The difference relations from \cite{IMMMfe,Anton,arthdiff} are also generalized --
to 5-graded polynomials ${\cal S}(A,q,t,\sigma,h)$:
\be
{\cal S}_{[r_1^s]} - {\cal S}_{[r_2^s]} \ \ {\rm is \ divisible \ by} \ \
h\{A\sigma/t^s\} \{Aq^{r_1+r_2}/\sigma\}
\nn\\
{\cal S}_{[r^{s_1}]} - {\cal S}_{[r^{s_2}]} \ \ {\rm is \ divisible \ by} \ \
h\{Aq^r/\sigma\} \{A\sigma/t^{s_1+s_2}\}
\ee

Also, in the infinitesimal vicinity of the point $q=t=1$ we   have for our
rectangular superpolynomials:
\be
\left.
\frac{d(H_R - H_{[1]}^{|R|})}{dq}
\right|_{q = t = 1} = \nu_R \sigma_1^{|R|-2} \sigma_2 \nn \\
\left.\frac{d(H_R - H_{[1]}^{|R|})}{dt}
\right|_{q = t = 1} = -\nu_{R^t} \sigma_1^{|R|-2} \sigma_2,
\label{infiexpan}
\ee
where
\be
\nu_\lambda = \sum_i \lambda_i (\lambda_i-1),
\ee
and $\sigma_1, \sigma_2$ are the lowest special polynomials.
In particular,
\be
\left.
\frac{d(H_{[1^s]} - H_{[1]}^{s})}{dq}
\right|_{q = t = 1} = 0 \nn \\
\left.\frac{d(H_{[r]} - H_{[1]}^{r})}{dt}
\right|_{q = t = 1} = 0,
\ee
Eqs.(\ref{infiexpan}) validate the conjecture (15) of \cite{Anton2}
for all rectangular representations $R$, at least in the case of the figure-eight knot.

\section{Conclusion}

In this paper we reported a substantial new space in the construction of
colored superpolynomials: the answer (\ref{41rssup}) is suggested for the figure-eight knot,
which is non-torus.
The suggestion is for all rectangular representations $R=[r^s]$
and it is always positive.
It reproduces the previous suggestions \cite{DGR,IMMMfe,supA,GGS,NO}
for symmetric and antisymmetric representations $[r]$ and $[1^s]$ --
on which there seems to be a consensus in the literature
(with the single exception of \cite{ArthSha}).

The answer (\ref{41rsmod}) is an immediate (most naive) deformation of the
recently suggested \cite{rect41} formula for rectangular HOMFLY --
after it is rewritten in the elegant form (\ref{41rsmod}),
which is another achievement of this paper.
It has an amusingly suggestive structure -- which, however, is specific for
rectangular representations and  remains to be better understood.

Further generalizations to non-rectangular representations and other knots
should follow.
Especially hopeful is the situation with twisted knots in rectangular
representations, where HOMFLY was recently found in \cite{rectwist}.
All these suggestions are based on the study of differential expansion
\cite{IMMMfe,evo,arthdiff,Konodef}, which strongly depends on the {\it defect}
of the knot \cite{Konodef}.
Thus the natural sequence of steps would be to first look at twisted knots,
then at other knots with defects zero and minus one -- and then proceed to
knots with positive defects, including the torus ones, for which the {\it hyper}polynomials
were suggested by I.Cherednik \cite{Che}.
These torus {\it hyper}polynomials are also positive in rectangular representations
and thus have good chances to be rectangular {\it super}polynomials -- thus they
can be compared with the implications of differential expansion,
when they will be found.
There is, however, a considerable work to do in this direction.


\section*{Acknowledgements}

Our work is partly supported by grants RFBR grants 16-01-00291 (Y.K.), 16-02-01021 (A.M.)
by young scientist grants 16-31-00484 (Y.K.), 15-31-20832-mol-a-ved (A.M.),
by Simons Foundation (Y.K.) and by the joint grants
15-51-52031-HHC, 15-52-50041-YaF, 16-51-53034-GFEN, 16-51-45029-Ind.


\newpage

\begin{thebibliography}{12}


\bibitem{CS} S.-S. Chern and J. Simons,
Ann.Math. {\bf 99} (1974) 48-69\\
E. Witten,
Comm.Math.Phys. {\bf 121} (1989)  351-399

\bibitem{knotpols}
J.W.Alexander, Trans.Amer.Math.Soc. \textbf{30} (2) (1928) 275-306 \\
J.H.Conway, Algebraic Properties, In: John Leech (ed.), Computational Problems in Abstract Algebra, Proc. \\
Conf. Oxford, 1967, Pergamon Press, Oxford-New York, 329-358, 1970 \\
V.F.R.Jones, Invent.Math. \textbf{72} (1983) 1; Bull.AMS \textbf{12} (1985) 103; Ann.Math. \textbf{126} (1987) 335 \\
L.Kauffman,Topology \textbf{26} (1987) 395 \\
P.Freyd, D.Yetter, J.Hoste, W.B.R.Lickorish, K.Millet, A.Ocneanu, Bull. AMS. \textbf{12} (1985) 239 \\
J.H.Przytycki and K.P.Traczyk, Kobe J. Math. \textbf{4} (1987) 115-139 \\
E.Witten, E.Witten, Comm.Math.Phys. {\bf 121} (1989) 351\\
A.Morozov and A.Smirnov, arXiv:1307.2576




\bibitem{GSV}
R.Gopakumar and C.Vafa hep-th/9802016, hep-th/9809187, hep-th/9811131, hep-th/9812127\\
H. Ooguri and C. Vafa, Nucl.Phys. {\bf B577} (2000) 419-438, hep-th/9912123 \\
S.Gukov, A.Schwarz and C.Vafa, Lett.Math.Phys. \textbf{74} (2005) 53-74, hep-th/0412243 \\
M.Dedushenko and E.Witten, arXiv:1411.7108


\bibitem{Kho} M.Khovanov, Duke Math.J. \textbf{101} (2000) no.3, 359426, math/9908171

\bibitem{KhR} M.Khovanov and L.Rozansky, Fund. Math. \textbf{199} (2008), no. 1, \textbf{191}, math/0401268;
Geom.Topol. \textbf{12} (2008), no.
3, 13871425, math/0505056; math/0701333 \\
N.Carqueville and D.Murfet, arXiv:1108.1081


\bibitem{padic} A.Morozov, Theor.Math.Phys. \textbf{187} (2016) no.1, 447-454, arXiv:1509.04928

\bibitem{GGS} E.Gorsky, S.Gukov, M.Stosic, arXiv:1304.3481

\bibitem{arthdiff} S.Arthamonov, A.Mironov, A.Morozov, Theor.Math.Phys.
\textbf{179} (2014) 509-542 (Teor.Mat.Fiz. 179 (2014) 147-188), arXiv:1306.5682



\bibitem{Ok} H.Nakajima, {\it Lectures on Hilbert Schemes of Points on Surfaces};
Adv. Studies in Pure Math. \textbf{69}, 2016, Development of Moduli Theory -- Kyoto \textbf{2013}, 173-205\\
A.Okounkov, arXiv:1512.07363 \\
N.Nekrasov, A.Okounkov, arXiv:1404.2323 \\
E.Carlsson, N.Nekrasov, A.Okounkov, arXiv:1308.2465 \\
E.Carlsson, A.Okounkov, arXiv:0801.2565

\bibitem{DIM}
J. Ding, K. Iohara, Lett. Math. Phys. 41 (1997) 181–193, q-alg/9608002 \\
K. Miki, J. Math. Phys. 48 (2007) 123520 \\
B.Feigin, E.Feigin, M.Jimbo, T.Miwa, E.Mukhin, Kyoto J. Math. 51 (2011) 337-364; 365-392;
 arXiv:1002.3100; arXiv:1002.3113 \\
 A.Mironov, A.Morozov, Sh.Shakirov, A.Smirnov, Nucl.Phys. B 855 (2012)  128-151, arXiv:1105.0948 \\
 H. Awata, B. Feigin and J. Shiraishi, arXiv:1112.6074 \\
  D. Maulik and A. Okounkov, arXiv:1211.1287 \\
N.Nekrasov, V.Pestun and S.Shatashvili,  arXiv:1312.6689 \\
M.Bershtein and  O.Foda, JHEP 2014  (2014) 177, arXiv:1404.7075 \\
V.Belavin, O.Foda and R.Santachiara, JHEP 10 (2015) 073, arXiv:1507.03540  \\
A.Morozov and Y.Zenkevich,   JHEP 1602 (2016) 098, arXiv:1510.01896 \\
J.-E.Bourgine, Y.Matsuo and H.Zhang, arXiv:1512.02492 \\
N.Nekrasov, arXiv:1512.05388,  arXiv:1608.07272 \\
 T.Kimura and V.Pestun,   arXiv:1512.08533, arXiv:1608.04651 \\
 A.Mironov, A.Morozov, Y.Zenkevich,   JHEP  05 (2016) 1-44,  arXiv:1603.00304;
 Phys.Lett. (2016), arXiv:1603.05467 \\
 H.Awata, H.Kanno, T.Matsumoto, A.Mironov, A.Morozov, An.Morozov, Yu.Ohkubo and Y.Zenkevich,
 JHEP 07 (2016) 1-67,  arXiv:1604.08366;   arXiv:1608.05351 \\
 A.Nedelin, F.Nieri and M.Zabzine, arXiv:1605.07029 \\
J.-E.Bourgine, M.Fukuda, Y.Matsuo and H.Zhang, arXiv:1606.08020



\bibitem{DGR} N.Dunfield, S.Gukov, J.Rasmussen,
Experiment.Math. \textbf{15} (\textbf{2}) (2006) 129-160, math/0505662

\bibitem{AgSha}   M.Aganagic and Sh.Shakirov, arXiv:1105.5117, arXiv:1202.2489, arXiv:1210.2733

\bibitem{DMMSS}  P.Dunin-Barkowski, A.Mironov, A.Morozov, A.Sleptsov, A.Smirnov,
JHEP 03 (2013) 021, arXiv:1106.4305

\bibitem{Che} I.Cherednik, arXiv:1111.6195

\bibitem{HallLitt}
A.Mironov, A.Morozov, Sh.Shakirov, A.Sleptsov,
JHEP 2012 (2012) 70, arXiv:1201.3339 \\
A.Mironov, A.Morozov, Sh.Shakirov, J. Phys. A: Math. Theor. 45 (2012) 355202,
 arXiv:1203.0667

\bibitem{betadefo} A. Morozov, Theor.Math.Phys. 173 (2012) 1417-1437, arXiv:1201.4595

\bibitem{IMMMfe} H. Itoyama, A. Mironov, A. Morozov and An. Morozov,
JHEP {\bf 2012} (2012) 131, arXiv:1203.5978

\bibitem{FSS} H.Fuji, S.Gukov and P.Sulkowski, arXiv:1205.1515

\bibitem{Anton} Anton Morozov, JHEP 1212 (2012) 116, arXiv:1208.3544

\bibitem{supA} S.Nawata, P.Ramadevi, Zodinmawia, X.Sun, JHEP 1211 (2012) 157, arXiv:1209.1409

\bibitem{FuSu} H.Fuji and P.Sulkowski, Procs of String Math 2012, arXiv:1303.3709

\bibitem{Rama} S.Nawata, P. Ramadevi and Zodinmawia, JHEP {\bf 1401} (2014) 126, arXiv:1310.2240

\bibitem{MMSsup}  A. Mironov, A. Morozov, A. Sleptsov, A. Smirnov,
Nucl.Phys. B889 (2014) 757-777, arXiv:1310.7622

\bibitem{Berest} Yu.Berest and P.Samuelson, Compositio Mathematica 152 (2016) 1333-1384, arXiv:1402.6032

\bibitem{CheDa} I.Cherednik and I.Danilenko, arXiv:1509.08351, arXiv:1408.4348

\bibitem{ArthSha} S.Arthamonov and Sh.Shakirov,  arXiv:1504.02620

\bibitem{Wit} E.Witten, arXiv:1603.03854

\bibitem{GNSSS} S. Gukov, S. Nawata, I. Saberi, M. Stosic and P. Sulkowski, arXiv:1512.07883

\bibitem{NO}  S.Nawata and  A.Oblomkov, arXiv:1510.01795






\bibitem{Vog}
P. Vogel, The universal Lie algebra, preprint (1999), see at http://webusers.imj-prg.fr/˜pierre.vogel/;
J. Pure Appl. Algebra 215 (2011) 1292-1339 \\
P. Deligne, C.R.Acad.Sci. 322 (1996) 321326 \\
A. Cohen, R. de Man, C.R.Acad.Sci. 322 (1996) 427432 \\
J. M. Landsberg, and, L. Manivel, arXiv:math.AG/0203241; Adv.Math. 171 (2002) 59-85


\bibitem{unipols}
A.Mironov, R.Mkrtchyan, A.Morozov, JHEP 02 (2016) 78,  arXiv:1510.05884;
Phys.Lett. B755 (2016) 47-57, arXiv:1511.09077


\bibitem{mmmspre}
D.Galakhov, D.Melnikov, A.Mironov, A.Morozov, A.Sleptsov,
 Phys.Lett. B743 (2015) 71, arXiv:1412.2616  \\
A. Mironov, A. Morozov, A. Sleptsov, JHEP 07 (2015) 069,  arXiv:1412.8432







\bibitem{evo} A.Mironov, A.Morozov, An.Morozov, AIP Conf. Proc. \textbf{1562} (2013) 123, arXiv:1306.3197

\bibitem{arthlinks} S.Arthamonov, A.Mironov, A.Morozov, An.Morozov, JHEP 04 (2014) 156, arXiv:1309.7984

\bibitem{Konodef} Ya.Kononov and A.Morozov, JETP Letters 101 (2015) 831-834,  arXiv:1504.07146

\bibitem{rect41} A.Morozov, Nucl.Phys. (2016), arXiv:1605.09728


\bibitem{hcube} D.Bar-Natan, Algebraic and Geometric Topology \textbf{2} (2002) 337-370, math/0201043 \\
V.Dolotin and A.Morozov, Nucl.Phys. \textbf{B878} (2014) 12-81, arXiv:1308.5759 \\
A.Morozov, An.Morozov, A.Popolitov, Phys.Lett.B749 (2015) 309-325,  arXiv:1506.07516; arXiv:1508.01957

\bibitem{Anosup} A.Anokhina and A.Morozov, JHEP \textbf{07} (2014) 063, arXiv:1403.8087



\bibitem{RT}
N.Yu.Reshetikhin and V.G.Turaev, Comm. Math. Phys. \textbf{127} (1990) 1-26 \\
E.Guadagnini, M.Martellini and M.Mintchev, Clausthal 1989, Procs.307-317;
Phys.Lett. \textbf{B235} (1990) 275 \\
V. G. Turaev and O. Y. Viro, Topology \textbf{31}, 865 (1992) \\
A.Morozov and A.Smirnov, Nucl.Phys. \textbf{B835} (2010) 284-313, arXiv:1001.2003 \\
A.Smirnov, Proc. of International School of Subnuclar Phys. Erice, Italy, 2009, arXiv:hep-th/0910.5011

\bibitem{inds}
R.K.Kaul, T.R.Govindarajan, Nucl.Phys. B380 (1992) 293-336, hep-th/9111063; B393 (1993) 392-412 \\
P.Ramadevi, T.R.Govindarajan and R.K.Kaul, Nucl.Phys. B402 (1993) 548-566, hep-th/9212110 \\
Nucl.Phys. B422 (1994) 291-306, hep-th/9312215; Mod.Phys.Lett. A10 (1995) 1635-1658, hep-th/9412084
P.Ramadevi and T.Sarkar, Nucl.Phys. B600 (2001) 487-511, hep-th/0009188 \\
P.Ramadevi and Zodinmawia, arXiv:1107.3918; arXiv:1209.1346 \\
S.Nawata, P.Ramadevi, Zodinmawia, J.Knot Theory and Its Ramifications 22 (2013) 13, arXiv:1302.5144 \\
Zodinmawia's PhD thesis, 2014

\bibitem{RTmod} A.Mironov, A.Morozov, An.Morozov, in Memorial Volume for Max Kreuzer,
 arXiv:1112.5754;   JHEP 03 (2012) 034, arXiv:1112.2654 \\
  S.Nawata, P.Ramadevi and Vivek Kumar Singh, arXiv:1504.00364 \\
 A. Mironov, A. Morozov, An. Morozov, P. Ramadevi, Vivek Kumar Singh, A. Sleptsov,
  JHEP 1507 (2015) 109,  arXiv:1504.00371;   arXiv:1601.04199 \\
A. Mironov, A. Morozov, Nucl.Phys. B899 (2015) 395-413,   arXiv:1506.00339


\bibitem{KaufR}  L.Kauffman, Topology 26 (1987) 395-407; Trans.Amer.Math.Soc. 311 (1989) 697-710 \\
L.Kauffman and P.Vogel, J.Knot Theory Ramifications 1 (1992) 59-104




\bibitem{AnoM} A. Anokhina and An. Morozov, Theor.Math.Phys. 178 (2014) 1-58,  arXiv:1307.2216




\bibitem{GaMoore} D.Galakhov and G.Moore, arXiv:1607.04222


\bibitem{lfp} P.Griffiths, J.Harris,
{\it Principles of algebraic geometry}, Wiley Classics Library, New York, 1994



\bibitem{rectwist}
 A.Morozov, arXiv:1606.06015




\bibitem{Ano21} A. Anokhina, A. Mironov, A. Morozov, An. Morozov,
Nucl.Phys. B \textbf{882C} (2014) 171-194,     arXiv:1211.6375

\bibitem{MMMS21}  A.Mironov, A.Morozov, An.Morozov, A.Sleptsov,
Int.J.Mod.Phys. A30 (2015) 1550169, arXiv:1508.02870

\bibitem{mmms1} A. Mironov, A. Morozov, An. Morozov, A. Sleptsov, arXiv:1605.02313

\bibitem{spepo}
Kefeng Liu and Pan Peng, arXiv:0704.1526 \\
Shengmao Zhu, arXiv:1206.5886

\bibitem{KM} Ya.Kononov and A.Morozov,  Phys.Lett. B747 (2015) 500-510, arXiv:1505.06170

\bibitem{Anton2} Anton Morozov,  JETP Lett. 97 (2013) 171-172, arXiv:1211.4596

\end{thebibliography}
\end{document}